\documentclass[letterpaper, 10 pt, conference]{ieeeconf}  

\IEEEoverridecommandlockouts                              

\usepackage{graphicx} 
\usepackage{amsmath} 
\usepackage{amssymb}  
\usepackage{balance}
\usepackage{url}

\title{\LARGE \bf
A Sampling Complexity-aware Framework for Discrete-time Fractional-Order Dynamical System Identification 
}

\author{Xiaole Zhang$^{1}$, Vijay Gupta$^{2}$, Paul Bogdan$^{1}$
\thanks{$^{1}$Ming Hsieh Department of Electrical and Computer Engineering, University of Southern California, Los Angeles, CA 90089, USA {\tt\small \{xiaolezh,pbogdan\}@usc.edu}}%
\thanks{$^{2}$Elmore Family School of Electrical and Computer Engineering, Purdue University, West Lafayette, IN 47907, USA {\tt\small gupta869@purdue.edu}}%
}

\begin{document}

\maketitle
\thispagestyle{empty}
\pagestyle{empty}

\begin{abstract}

A variety of complex biological, natural and man-made systems exhibit non-Markovian dynamics that can be modeled through fractional order differential equations, yet, we lack sample comlexity aware system  identification strategies. Towards this end, we propose an affine discrete-time fractional order dynamical system (FoDS) identification algorithm and provide a detailed sample complexity analysis. The algorithm effectively addresses the challenges of FoDS identification in the presence of noisy data. The proposed algorithm consists of two key steps. Firstly, it avoids solving higher-order polynomial equations, which would otherwise result in multiple potential solutions for the fractional orders. Secondly, the identification problem is reformulated as a least squares estimation, allowing us to infer the system parameters. We derive the expectation and probabilistic bounds for the FoDS parameter estimation error, assuming prior knowledge of the functions \( f \) and \( g \) in the FoDS model. The error decays at a rate of \( N = O\left( \frac{d}{\epsilon} \right) \), where \( N \) is the number of samples, \( d \) is the dimension of the state variable, and \( \epsilon \) represents the desired estimation accuracy. Simulation results demonstrate that our theoretical bounds are tight, validating the accuracy and robustness of this algorithm.

\end{abstract}

\section{INTRODUCTION}


A wide variety of biological (e.g., microscopic neuronal networks~\cite{lundstrom2008fractional,yin2020network} and macroscopic brain-region networks~\cite{gupta2018dealing,GuptaACC2019}, heart rate variability~\cite{ivanov1999multifractality,ivanov2003long,ivanov2021new}, blood glucose variability~\cite{ghorbani2014reducing}, gene regulatory networks~\cite{Ghorbani2018}), technological (e.g., network traffic~\cite{willinger2003long}, power grids~\cite{SiaCDC2018}, cyber-physical systems~\cite{XueICCPS2017}, brain-machine-body interfaces~\cite{XueDATE2016}) and social~\cite{belletti2019quantifying} systems have been demonstrated to exhibit long-range memory / dependence (LRM/LRD), fractal and non-Markovian dynamics. Unlike the Markovian or the short-range memory dynamical systems which are characterized by the fact that the current state only depends on the previous state and there is no memory effect from the historical trajectory of the dynamical systems, the non-Markovian dynamical systems exhibit a nontrivial dependence between the current and future states on a wide range of  past values / states. Simply speaking, the non-Markovian dynamics cannot be approximated by linear time-invariant systems. Despite an apparent setback, fractional calculus~\cite{hilfer2000applications,nonnenmacher2000applications,chen2009fractional,west2014colloquium,ionescu2017role,sun2018new,reed2022fractional} with its fractional-order derivative and integral concepts allow us to construct compact dynamical models (i.e., capture the dependence on a wide range of past values through just one parameter of the fractional operator) of complex systems. For example, fractional calculus concepts have been exploited for proposing feedback control laws for linear discrete-time fractional-order systems with additive state disturbance~\cite{alessandretti2020finite}, more efficient reinforcement learning strategies that account for non-Markovian dynamics~\cite{gupta2021non} or control stratgies for non-Markovian dynamical networks~\cite{reed2023role} that require orders of magnitude less number of nodes to be controlled than their Markovian counterparts. These efforts highlight the need for accurate inference of the parameters of the fractional-order dynamical systems (FoDS)~\cite{chatterjee2022learning,yaghooti2023inferring}.

In the context of integer-order systems, the problem of system identification becomes more complex when the order is unknown and must be estimated alongside the system parameters. For subspace based methods, Shibata~\cite{shibata1976selection} and Bauer~\cite{bauer2001order} address order estimation by analyzing information from estimated singular values and innovation variance. Ljung et al.~\cite{ljung2015regularization} propose identifying systems with unknown order by first learning the largest possible model that fits the data and then performing model reduction. Sarkar et al.~\cite{sarkar2021finite} address the problem of learning the parameters of a stable linear time-invariant (LTI) system with unkown order, focusing on determining the best lower-order approximation using finite data. He and Asada~\cite{he_asada_1993} propose a method for identifying the orders of input-output models for unknown nonlinear dynamic systems based on the continuity property of the nonlinear functions, specifically, using the Lipschitz quotient. Chen and Zhang~\cite{chen_zhang_1990} provide recursive estimates for the time-delay, system orders, and coefficients of single-input single-output linear discrete-time systems, given a lower bound for the time-delay and upper bounds for system orders.

While significant progress has been made in system identification for integer-order systems, the identification of FoDSs presents additional challenges. Yaghooti~\cite{yaghooti2023inferring} recently proposed a system identification algorithm for discrete fractional-order nonlinear systems. This algorithm consists of two main steps. It first introduces a data generation mechanism and then uses the generated data to estimate the parameters. The fractional order is identified by solving quadratic equations and then substituting the solutions into cubic equations to select the one that satisfies them. However, the algorithm fails to perform well under noisy data, particularly in high-dimensional settings. This is due to the impracticality of solving quadratic equations and verifying solutions against cubic polynomials when noise is present. To address this limitation, we propose a new FoDS identification algorithm capable of handling noisy data.

The main contributions of the present paper are as follows:
\textit{(i)} We propose a robust affine discrete-time fractional-order dynamical system identification algorithm that can handle unknown system orders. The algorithm effectively identifies the system even when the data is noisy.

\textit{(ii)} We derive expectation and probabilistic bounds for the fractional-order estimation error, assuming prior knowledge of the functions $f$ and $g$ in the system. The error decays at a rate of $N = O(d/\epsilon)$.

\section{PROBLEM FORMULATION}

In this section, we begin with the definition of the Gr\"{u}nwald-Letnikov operator. Next, we describe how to represent our discrete-time FoDS using the Gr\"{u}nwald-Letnikov difference operator. Finally, we state our system identification problem along with some assumptions.

\subsection{Gr\"{u}nwald-Letnikov Operator}

The fractional-order Grünwald–Letnikov operator allows to discretize the fractional order derivative and represent it as a finite difference as follows:
\begin{equation}
\Delta^\alpha x_k := \sum_{j=0}^{k} \psi(\alpha, j) x_{k-j},
\end{equation}
where \( x_k \in \mathbb{R}^d \), \( \alpha = [\alpha_1, \alpha_2, \ldots, \alpha_d]^\top \in \mathbb{R}^d \) represents the order of the difference operator, and \( \psi(\alpha, j) \in \mathbb{R}^{d \times d} \) is an \( d \times d \) diagonal matrix defined as
\begin{equation}
\psi(\alpha, j) := \operatorname{diag}(\psi(\alpha_1, j), \psi(\alpha_2, j), \ldots, \psi(\alpha_d, j)).
\end{equation}
with
\begin{equation}
\psi(\alpha_i, j) := \frac{\Gamma(j - \alpha_i)}{\Gamma(-\alpha_i) \Gamma(j + 1)}, \quad i = 1, 2, \ldots, d,
\end{equation}
where \( \Gamma(\cdot) \) denotes the Gamma function.

\subsection{Discrete-time Fractional-Order Dynamical System}

The state-space representation of the discrete-time FoDS reads:
\begin{equation}
\Delta^\alpha x_{k+1} = h(x_k, u_k),
\end{equation}
where \( x_k \in \mathbb{R}^d \) is the state vector, \( u_k \in \mathbb{R}^m \) is the system input, and \( h(\cdot, \cdot) = \begin{bmatrix} h_1(\cdot, \cdot) & \cdots & h_d(\cdot, \cdot) \end{bmatrix}^\top \). The functions \( h_i(\cdot, \cdot) : \mathbb{R}^d \times \mathbb{R}^m \to \mathbb{R} \), for \( i = 1, 2, \ldots, d \), can be any functions.

Using the Grünwald–Letnikov difference operator, we can write the system as follows:
\begin{equation}
x_{k+1} = h(x_k, u_k) - \sum_{j=1}^{k+1} \psi(\alpha, j) x_{k+1-j}.
\end{equation}
The second term in eq. (5) highlights the memory-dependent nature of FoDSs, making them suitable for modeling non-Markovian and long-range dependent processes.

\subsection{FoDS Model Estimation}

We aim to estimate the parameters of the following affine, discrete-time FoDS:
\begin{equation}
x_{k+1} = f(x_k) + g(x_k) u_k - \sum_{j=1}^{k+1} \psi(\alpha, j) x_{k+1-j}.
\end{equation}
Where \( x_k \in \Omega \subseteq \mathbb{R}^d \) is the state vector, and the functions \( f(\cdot): \Omega \to \mathbb{R}^d \) and \( g(\cdot): \Omega \to \mathbb{R}^{d \times m} \) are defined as:
\begin{equation}
f(\cdot) = \begin{bmatrix} f_1(\cdot) \\ \vdots \\ f_d(\cdot) \end{bmatrix},
\quad
g(\cdot) = \begin{bmatrix}
g_{11}(\cdot) & \cdots & g_{1m}(\cdot) \\
\vdots & \ddots & \vdots \\
g_{d1}(\cdot) & \cdots & g_{dm}(\cdot)
\end{bmatrix}.
\end{equation}

\textit{Problem 1.} Assume we have full knowledge of the functions \( f(\cdot) \) and \( g(\cdot) \), and that the measurements are subject to white Gaussian noise. Specifically, the initial state \( x_0 \) is noise-free, but the noise begins with \( x_1 \). Develop a system identification algorithm to estimate the fractional order \( \alpha \). Do a sample complexity analysis for the proposed algorithm. 

\textit{Problem 2.} Assume we have no knowledge of the functions \( f(\cdot) \) and \( g(\cdot) \), and that the measurements are subject to white Gaussian noise. Specifically, the initial state \( x_0 \) is noise-free, but the noise begins with \( x_1 \). Develop a system identification algorithm to estimate the fractional order \( \alpha \), functions \( f(\cdot) \) and \( g(\cdot) \). Perform a sample complexity analysis for the proposed algorithm.

Notice that this problem extends the prior work in \cite{yaghooti2023inferring}. In that paper, the authors proposed a system identification algorithm but did not consider the case where the data is noisy. They also did not analyze the sample complexity of their proposed algorithm.

\section{Parameter Estimation}

In this section, we propose a system identification algorithm to estimate the unknown parameters for Problem 2. It is evident that Problem 2 is a generalization of Problem 1. The proposed algorithm consists of two main steps. In the first step, experiments are designed to generate the necessary datasets. In the second step, the generated data from the first step is used to estimate the parameters by solving a least squares problem.

\subsection{Data Generation}

In this step, we generate data following the procedure outlined below:
\begin{align}
x_1^1 &= f(x_0^1) + g(x_0^1) u_0^1 - \psi(\alpha,1) x_0^1 + w_0^1 \\
&= f(x_0^1) + g(x_0^1) u_0^1 + A x_0^1 + w_0^1 \\
x_1^2 &= f(x_0^2) + g(x_0^2) u_0^2 - \psi(\alpha,1) x_0^2 + w_0^2 \\
&= f(x_0^2) + g(x_0^2) u_0^2 + A x_0^2 + w_0^2 \\
&\notag\text{...} \\
x_1^p &= f(x_0^p) + g(x_0^p) u_0^p - \psi(\alpha,1) x_0^p + w_0^p \\
&= f(x_0^p) + g(x_0^p) u_0^p + A x_0^p + w_0^p
\end{align}
where \( x_0^i \) are the \( i \)-th initial conditions, \( x_1^i \) represents the corresponding state at the second time step generated from the initial condition, and \( w_0^i \) are independent and identically distributed (i.i.d.) white Gaussian noise for \( i = 1, \ldots, p \), and \( A = \operatorname{diag}(\alpha_1, \alpha_2, \ldots, \alpha_d) \).

\subsection{Function Approximation}

The general approach for learning the unknown functions \( f(\cdot) \) and \( g(\cdot) \) is that we select two different sets of orthogonal basis functions and then optimize a least squares problem to derive the unknown coefficients for basis functions of \( f(\cdot) \) and \( g(\cdot) \). Without loss of generality, we illustrate the method for \( m = 1 \). Extending this method to \( m > 1 \) is straightforward.

To explain the case when \( m = 1 \), where the system input is \( u_k \in \mathbb{R} \) and \( g(\cdot) = [g_1(\cdot), \ldots, g_d(\cdot)]^T \), we approximate \( f_i(\cdot) \) and \( g_i(\cdot) \) using orthogonal basis functions. Specifically,
\begin{align}
f_i(x_k) &= \sum_{h=1}^{H} \gamma_i^{(h)} \pi_i^{(h)} (x_k), \\
g_i(x_k) &= \sum_{l=1}^{L} \beta_i^{(l)} \phi_i^{(l)} (x_k),
\end{align}
where \( \pi_i^{(h)}(\cdot): \Omega \to \mathbb{R} \) are orthogonal basis functions in the function space \( \mathcal{F} \) given \( f_i \in \mathcal{F} \), \( \phi_i^{(l)}(\cdot): \Omega \to \mathbb{R} \) are orthogonal basis functions in the space \( \mathcal{G} \) given \( g_i \in \mathcal{G} \), and \( \gamma_i^{(h)} \) and \( \beta_i^{(l)} \) are unknown coefficients. \( H \) and \( L \) denote the number of terms in the basis functions.

Then We have the following expressions:

\begin{align}
f(x_0^i) &= \begin{bmatrix}
\sum_{h=1}^{H} \gamma_1^{(h)} \pi_1^{(h)} (x_0^i) \\
\vdots \\
\sum_{h=1}^{H} \gamma_d^{(h)} \pi_d^{(h)} (x_0^i)
\end{bmatrix}, \\
g(x_0^i) u_0^i &= \begin{bmatrix}
\sum_{l=1}^{L} \beta_1^{(l)} \phi_1^{(l)} (x_0^i) \\
\vdots \\
\sum_{l=1}^{L} \beta_d^{(l)} \phi_d^{(l)} (x_0^i)
\end{bmatrix} u_0^i,
\end{align}
where \( i = 1, \ldots, p \).

\subsection{System Identification with Unknown System Order}

Using the function approximation technique, the original problem can be reformulated into a standard least squares problem. Without loss of generality, we explain this approach using the case where the system input dimension \( m = 1 \).

Let us define the following notation:
\[
\Omega =
\begin{bmatrix}
\operatorname{diag}(x_0^1) \\
\vdots \\
\operatorname{diag}(x_0^p)
\end{bmatrix}, \quad
\pi^i = 
\begin{bmatrix}
\pi_1^i & 0 & \cdots & \cdots \\
0 & \pi_2^i & 0 & \cdots \\
\vdots & \vdots & \ddots & \vdots\\
0 & \cdots & \cdots & \pi_d^i
\end{bmatrix},
\]

\[
\phi^i = 
\begin{bmatrix}
\phi_1^i & 0 & \cdots & \cdots \\
0 & \phi_2^i & 0 & \cdots \\
\vdots & \vdots & \ddots & \vdots\\
0 & \cdots & \cdots & \phi_d^i
\end{bmatrix} u_0^i,
\]
\[
\pi = \begin{bmatrix} \pi^1 \\ \vdots \\ \pi^p \end{bmatrix}, \quad \phi = \begin{bmatrix} \phi^1 \\ \vdots \\ \phi^p \end{bmatrix}, \quad X = \begin{bmatrix} x_1^1 \\ \vdots \\ x_1^p \end{bmatrix}, \quad w = \begin{bmatrix} w_0^1 \\ \vdots \\ w_0^p \end{bmatrix},
\]
\begin{equation}
\gamma = \begin{bmatrix} \gamma_1 \\ \vdots \\ \gamma_d \end{bmatrix}, \quad \beta = \begin{bmatrix} \beta_1 \\ \vdots \\ \beta_d \end{bmatrix}, \quad \alpha = [ \alpha_1, \cdots, \alpha_d ]^T,
\end{equation}
where \(\operatorname{diag}(\cdot)\) operator takes a \(d\)-dimensional vector and diagonalizes it into a \(d \times d\) matrix, \[\pi_k^i = [\pi_k^{(1)}(x_0^i), \pi_k^{(2)}(x_0^i), \cdots, \pi_k^{(H)}(x_0^i)],\] 
\[\phi_k^i = [\phi_k^{(1)}(x_0^i), \phi_k^{(2)}(x_0^i), \cdots, \phi_k^{(L)}(x_0^i)],\]
\(\gamma_i = [\gamma_i^{(1)}, \gamma_i^{(2)}, \cdots, \gamma_i^{(H)}]^T\), 
\(\beta_i = [\beta_i^{(1)}, \beta_i^{(2)}, \cdots, \beta_i^{(L)}]^T\), 
for \(i = 1, \ldots, p\), and \(k = 1, \ldots, d\). To be clear, \(\Omega \in \mathbb{R}^{dp \times d}\), \(\pi \in \mathbb{R}^{dp \times dH}\), and \(\phi \in \mathbb{R}^{dp \times dL}\).

Thus, given the generated data, we solve the following least squares problem:
\begin{equation}
\begin{bmatrix}
x_1^1 \\
\vdots \\
x_1^p
\end{bmatrix}
=
\pi
\begin{bmatrix}
\gamma_1 \\
\vdots \\
\gamma_d
\end{bmatrix}
+
\phi
\begin{bmatrix}
\beta_1 \\
\vdots \\
\beta_d
\end{bmatrix}
+
\begin{bmatrix}
\operatorname{diag}(x_0^1) \\
\vdots \\
\operatorname{diag}(x_0^p)
\end{bmatrix}
\begin{bmatrix}
\alpha_1 \\
\vdots \\
\alpha_d
\end{bmatrix}
+
\begin{bmatrix} 
w_0^1 \\ 
\vdots \\ 
w_0^p 
\end{bmatrix},
\end{equation}
\begin{equation}
X = \pi\gamma + \phi\beta + \Omega\alpha + w = \begin{bmatrix} \pi & \phi & \Omega \end{bmatrix}\begin{bmatrix} \gamma \\ \beta \\ \alpha \end{bmatrix} + w.
\end{equation}
Then we have the following optimization problem
\begin{equation}
\hat{\theta} = \arg\min_{\theta} \|X - \xi \theta\|_2^2,
\end{equation}
with the solution
\begin{equation}
\hat{\theta} = \left(\xi^T \xi\right)^{-1} \xi^T X,
\end{equation}
where \( \theta = [\gamma, \beta, \alpha]^T\), and \( \xi = [\pi, \phi, \Omega]\).

The proposed algorithm can be straightforwardly generalized to address Problem 1. To achieve this, rearrange the terms involving \( f(\cdot) \) and \( g(\cdot) \) to the left side of the data generation equation, and then solve the least squares problem.

\section{Sample Complexity Analysis}

It is natural to ask about the number of samples required by the fractional-order system identification algorithm to ensure that the error is controlled below a specified threshold. In this paper, we provide both an expectation bound and a high-probability bound for Problem 1. The sample complexity analysis for Problem 2, which includes function approximation errors, will be addressed in future work.

Assuming we have full knowledge of function \( f(\cdot) \) and \( g(\cdot) \), we can rearrange the terms involving \( f(\cdot) \) and \( g(\cdot) \) to the left side of data generation equation: 
\begin{equation}
\begin{bmatrix}
y_{k1}^1 \\
\vdots \\
y_{k1}^p
\end{bmatrix}
=
\begin{bmatrix}
\operatorname{diag}(x_0^1) \\
\vdots \\
\operatorname{diag}(x_0^p)
\end{bmatrix}
\begin{bmatrix}
\alpha_1 \\
\vdots \\
\alpha_d
\end{bmatrix}
+
\begin{bmatrix} 
w_{k0}^1 \\ 
\vdots \\ 
w_{k0}^p 
\end{bmatrix},
\end{equation}
where \(y_{k1}^i = x_{k1}^i - [f(x_0^i) + g(x_0^i) u_0^i] \) for \(i = 1, \ldots, p\).

The above is a specific sample generated from a given initial condition. We aim to determine how many such samples are required to control the error below a specified threshold, given a fixed initial condition. Specifically, we want to determine the total number \( N \) of observations \( y_k \), where \( y_k = [y_{k1}^1, \cdots, y_{k1}^p]^T \), and \( k = 1, \dots, N \).

We can now express the state-space model related to the sample complexity problem as follows:
\begin{equation}
\beta_{k+1} = \beta_k,
\end{equation}
\begin{equation}
y_k = C\beta_k + w_k,
\end{equation}
where the state vector \( \beta_k = [\alpha_1, \dots, \alpha_d]^T \), the measurement matrix \( C = \begin{bmatrix} \operatorname{diag}(x_0^1) \\ \vdots \\ \operatorname{diag}(x_0^p) \end{bmatrix} \), and \( w_k \) is a white Gaussian noise vector, \( w_k \sim \mathcal{N}(0, K_w) \), where \( K_w \) is a diagonal covariance matrix.

Our solution to \( \hat{\beta} \) is given by:
\begin{equation}
\hat{\beta} = (C^T C)^{-1} C^T \bar{Y},
\end{equation}
which is the result of optimizing the following least squares problem:
\begin{equation}
\hat{\beta} = \arg \min_{\beta} \| Y - A \beta \|_2^2,
\end{equation}
where \( \bar{Y} = \frac{1}{N} \sum_{k=1}^{N} Y_k \), \( A = [C^T, C^T, \dots, C^T]^T \), and \( A \) contains \( N \) repetitions of the matrix \( C \).

\subsection{Expectation Bound}

\textit{Theorem 1. For the solution \(\hat{\beta} = (C^T C)^{-1} C^T \bar{Y}\), the following expectation bound holds:
\begin{equation}
\mathbb{E}\left[\|\hat{\beta} - \beta\|_2^2\right] \leq \frac{d}{N} \frac{\lambda_{\text{max}}(C^T C) \lambda_{\text{max}}(K_w)}{\lambda_{\text{min}}(C^T C)^2},
\end{equation}
where \( \lambda_{\text{max}} \) and \( \lambda_{\text{min}} \) denote the largest and smallest eigenvalues, respectively, and \( K_w \) is the covariance matrix of the noise.}

The detailed proof is provided in Appendix \uppercase\expandafter{\romannumeral 1\relax}. The key intuition behind the proof is as follows. To bound \(\mathbb{E}\left[\|\hat{\beta} - \beta\|_2^2\right] \), we first notice that the error, \( \hat{\beta} - \beta \), follows a Gaussian distribution 
\[ N\left(0, \frac{1}{N} (C^T C)^{-1} C^T K_w C (C^T C)^{-1}\right).
\]
Then we use trace trick and \textit{Lemma 1} to derive the expectation bound.\hfill $\square$

\textit{Theorem 1} shows that the sample complexity of the fractional-order system identification algorithm depends on the dimension of the fractional order, which is also the dimension of the state vector. Specifically, given a tolerance \(\epsilon\), the sample complexity \( N \) scales as \( N = O\left(\frac{d}{\epsilon}\right) \), with the constant factor depending on the noise covariance matrix \( K_w \) and the measurement matrix \( C \).

\subsection{High-probability Bound}

\textit{Theorem 2. For the solution \(\hat{\beta} = (C^T C)^{-1} C^T \bar{Y}\), the following high-probability bound holds for \( t > \frac{\lambda d}{2}\):
\begin{equation}
\mathbb{P}(\|\hat{\beta} - \beta\|_2^2 \geq t) \leq \exp\left(-\frac{\frac{t}{\lambda} + \sqrt{\frac{2td}{\lambda} - d^2}}{2}\right),
\end{equation}
where \(\lambda = \lambda_{\text{max}}(\frac{1}{N} (C^T C)^{-1} C^T K_w C (C^T C)^{-1})\).} 

The detailed proof is provided in Appendix \uppercase\expandafter{\romannumeral 2\relax}. The key intuition behind the proof is as follows. We begin by whitening the error covariance matrix, transforming the error into a multivariate standard Gaussian random variable. This allows us to express the squared estimation error, \( \|\hat{\beta} - \beta\|_2^2 \), as a weighted sum of independent, identically distributed standard normal random variables. The squared error can then be upper-bounded by using a chi-squared distribution with \(d\) degrees of freedom. Using known high-probability bounds \textit{Lemma 2} for chi-squared distributed random variables~\cite{laurent2000adaptive}, we derive the high-probability bound for our algorithm. \hfill $\square$

\textit{Theorem 2} shows that the estimation error decreases exponentially as the threshold \(t\) increases, which is typical for high-probability bounds based on concentration inequalities. The decay rate is influenced by the noise level \(K_w\), measurement matrix \(C\), and the fractional order dimension \(d\). When \( t \) is very small, the term \( \sqrt{\frac{2td}{\lambda} - d^2} \) becomes dominant, and the bound suggests that the probability of error exceeding \( t \) is relatively high. However, as \( t \) increases, the probability of having an error greater than \( t \) becomes exponentially small.

Note that we also provide an alternative high-probability bound derived from known results for sub-exponential random variables~\cite{wainwright2019high} in the Appendix. However, this bound is not as tight as the one presented in \textit{Theorem 2}.

\section{Simulation}

In this section, we present the simulation results for the proposed fractional-order system identification algorithm. The simulations cover both cases where the functions \( f \) and \( g \) are known and unknown. We also demonstrate the sample complexity analysis by showing how the estimation error decreases as the number of samples \( N \) increases.

\subsection{Model Parameters Estimation}

Consider the following affine discrete-time fractional-order system:
\begin{align}
\Delta^\alpha x_{k+1} &= \mu x_k (1 - x_k) \nonumber \\
&\quad + a \cos(x_k) \exp\left(b[\sin(x_k - c\pi) - 1]\right) u_k. \label{eq:system}
\end{align}

Substituting (25) into the system (6), we have: 
\begin{equation}
f(x_k) = \mu x_k (1 - x_k),    
\end{equation}
\begin{equation}
g(x_k) = a \cos(x_k) \exp\left(b[\sin(x_k - c\pi) - 1]\right).    
\end{equation}

In our simulations, we evaluate the algorithm's performance in two cases: when the functions \(f\) and \(g\) are known, and when they are unknown. We set the parameters $\mu = 1$, $a = -1$, $b = 4$, and $c = 0.7$. In the case where the expressions of \(f\) and \(g\) are not provided, we approximate them using orthogonal basis functions. Specifically, for the function \(f\), we use the following trigonometric basis:
\begin{equation}
f(x) = \left[\sin(\pi x), \cos(\pi x), \dots, \cos((H-1)\pi x)\right].
\end{equation}
For the function \(g\), we use Chebyshev polynomials, but we skip the traditional second term. The basis for \(g\) is given by:
\begin{equation}
g(x) = \left[T_0(x), T_2(x), T_3(x), \dots, T_{L}(x)\right],
\end{equation}
where \(T_j(x)\) denotes the \(j\)-th Chebyshev polynomial. The drop of the second term is necessary because when the dimension of the state vector is one, \(\operatorname{diag}(x_k)\) simplifies to \(x_k\). Thus involving the second term will lead to a matrix non-invertibility issue. 
\begin{figure}[htbp]
    \centering
    \includegraphics[width=0.48\textwidth]{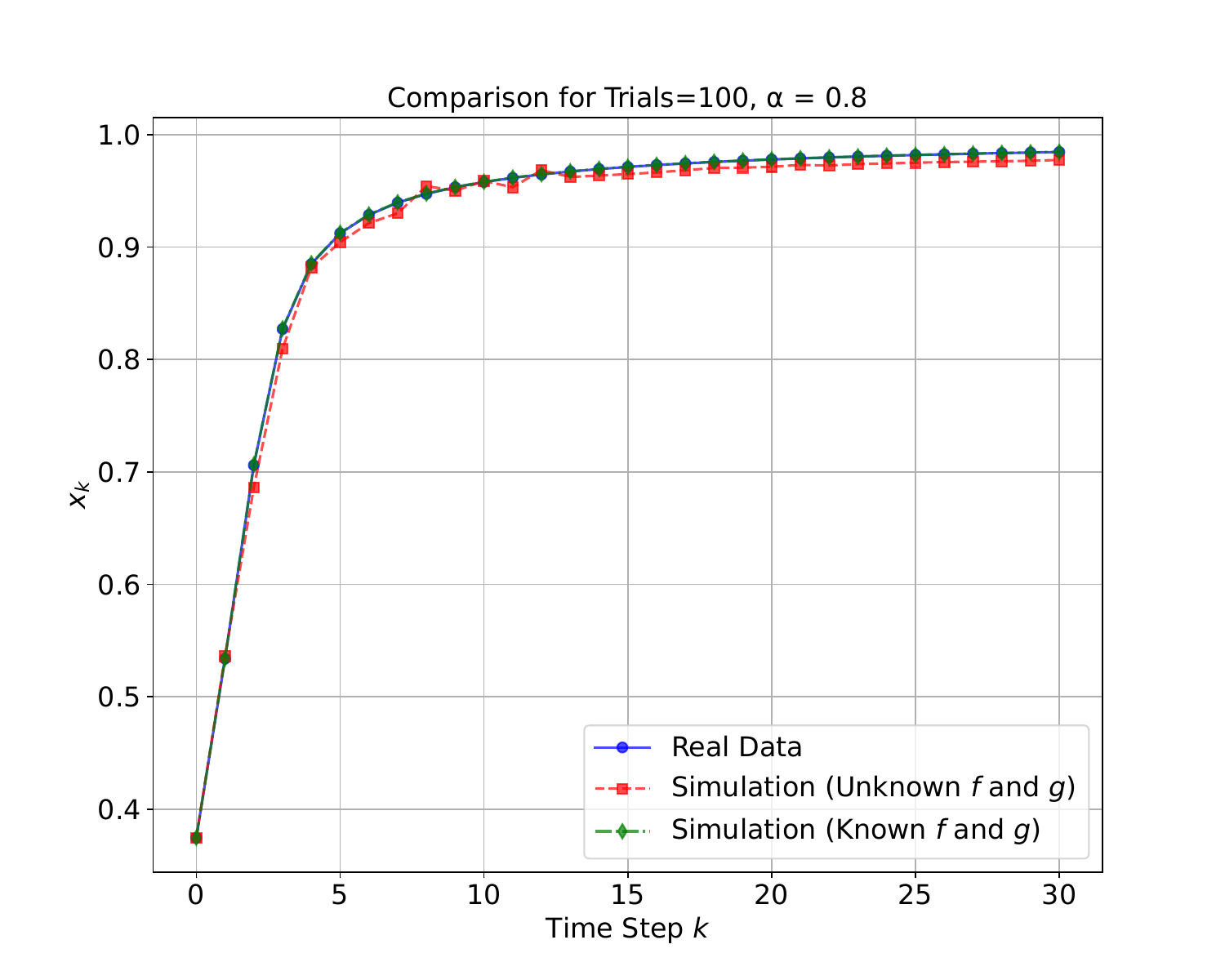}
    \caption{Comparison of simulations with 100 trials $\alpha = 0.8$.}
    \label{fig:trials_100_1}
\end{figure}

\begin{figure}[htbp]
    \centering
    \includegraphics[width=0.48\textwidth]{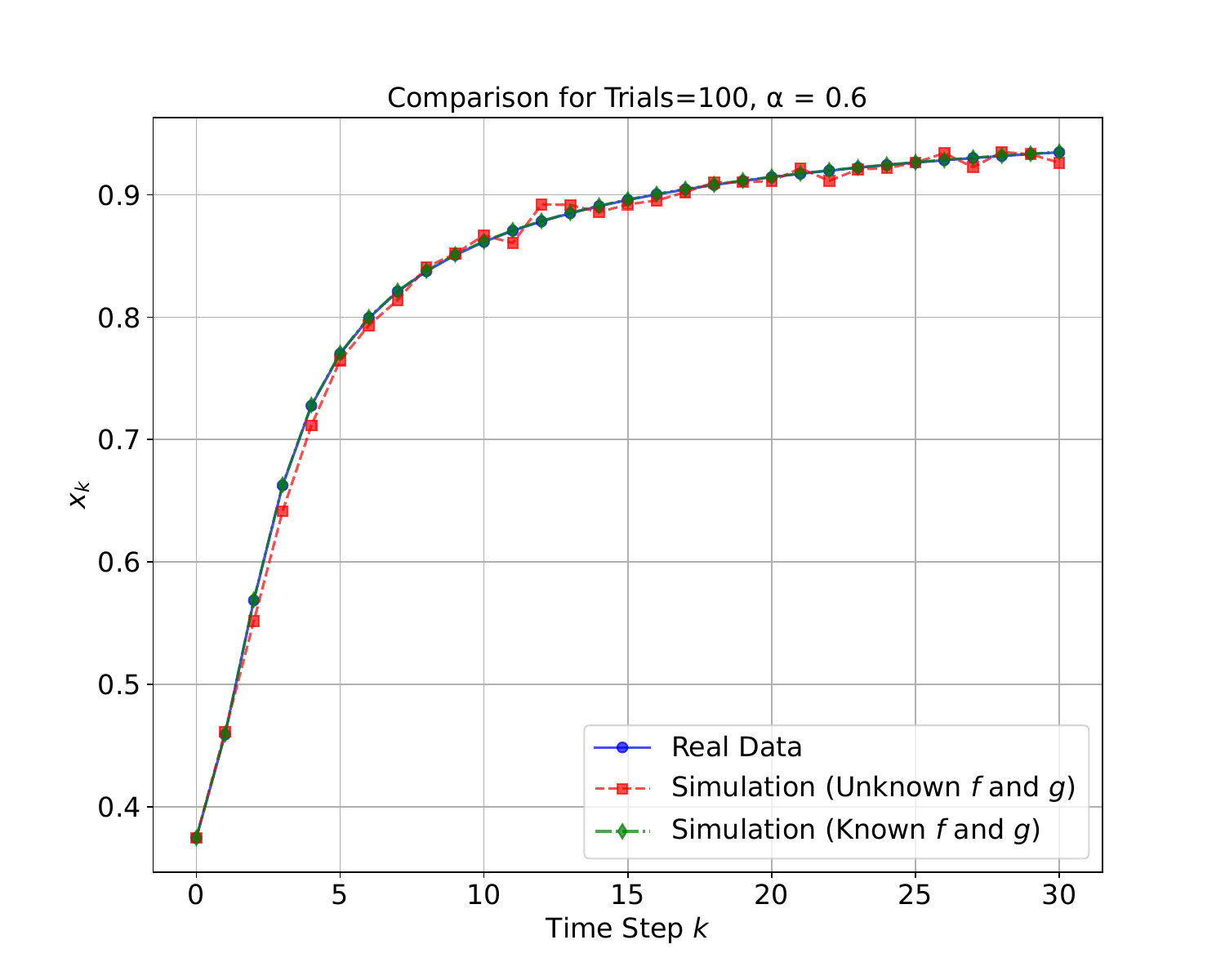}
    \caption{Comparison of simulations with 100 trials $\alpha = 0.6$.}
    \label{fig:trials_100_2}
\end{figure}

\begin{figure}[htbp]
    \centering
    \includegraphics[width=0.48\textwidth]{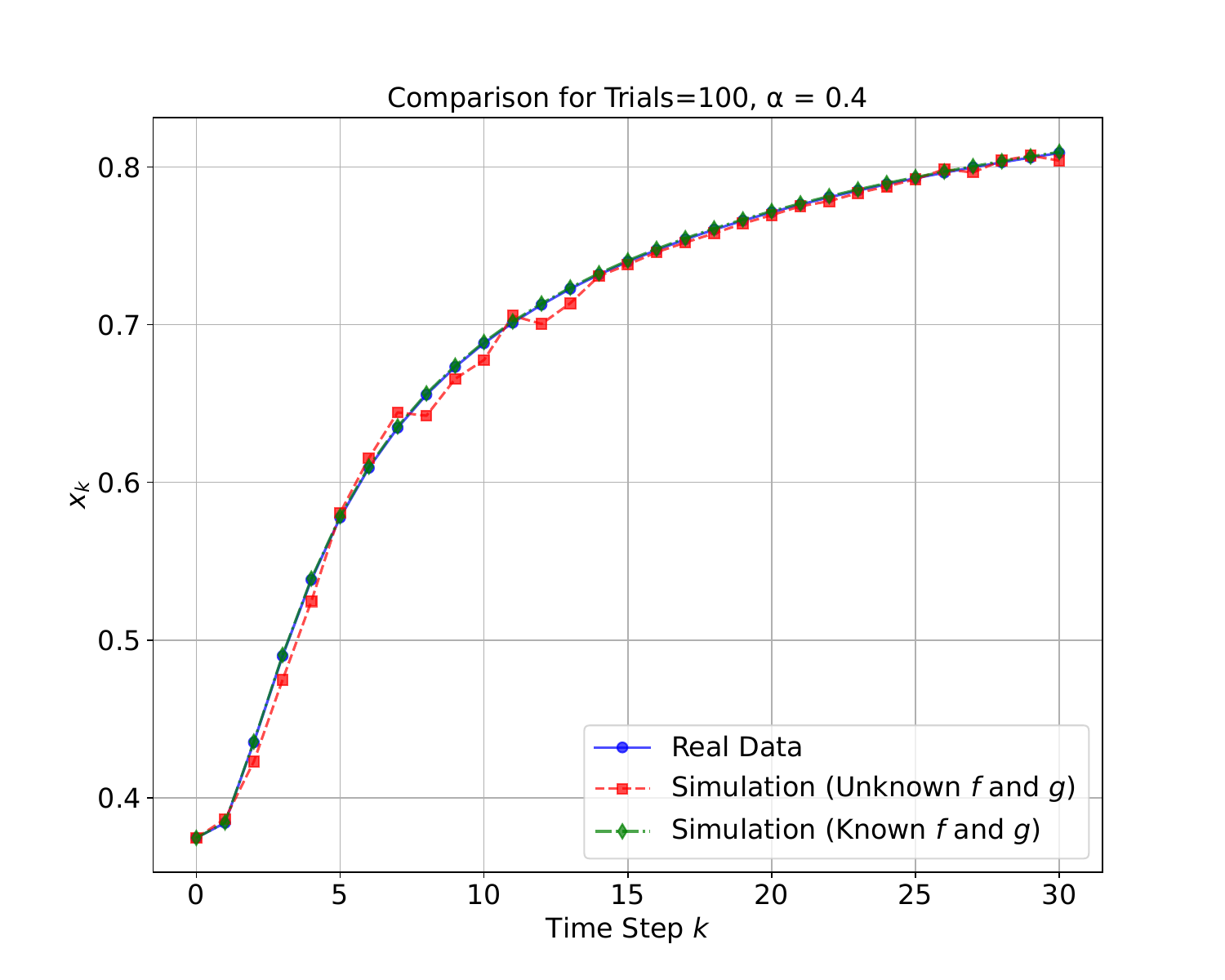}
    \caption{Comparison of simulations with 100 trials $\alpha = 0.4$.}
    \label{fig:trials_100_3}
\end{figure}

\begin{figure}[htbp]
    \centering
    \includegraphics[width=0.48\textwidth]{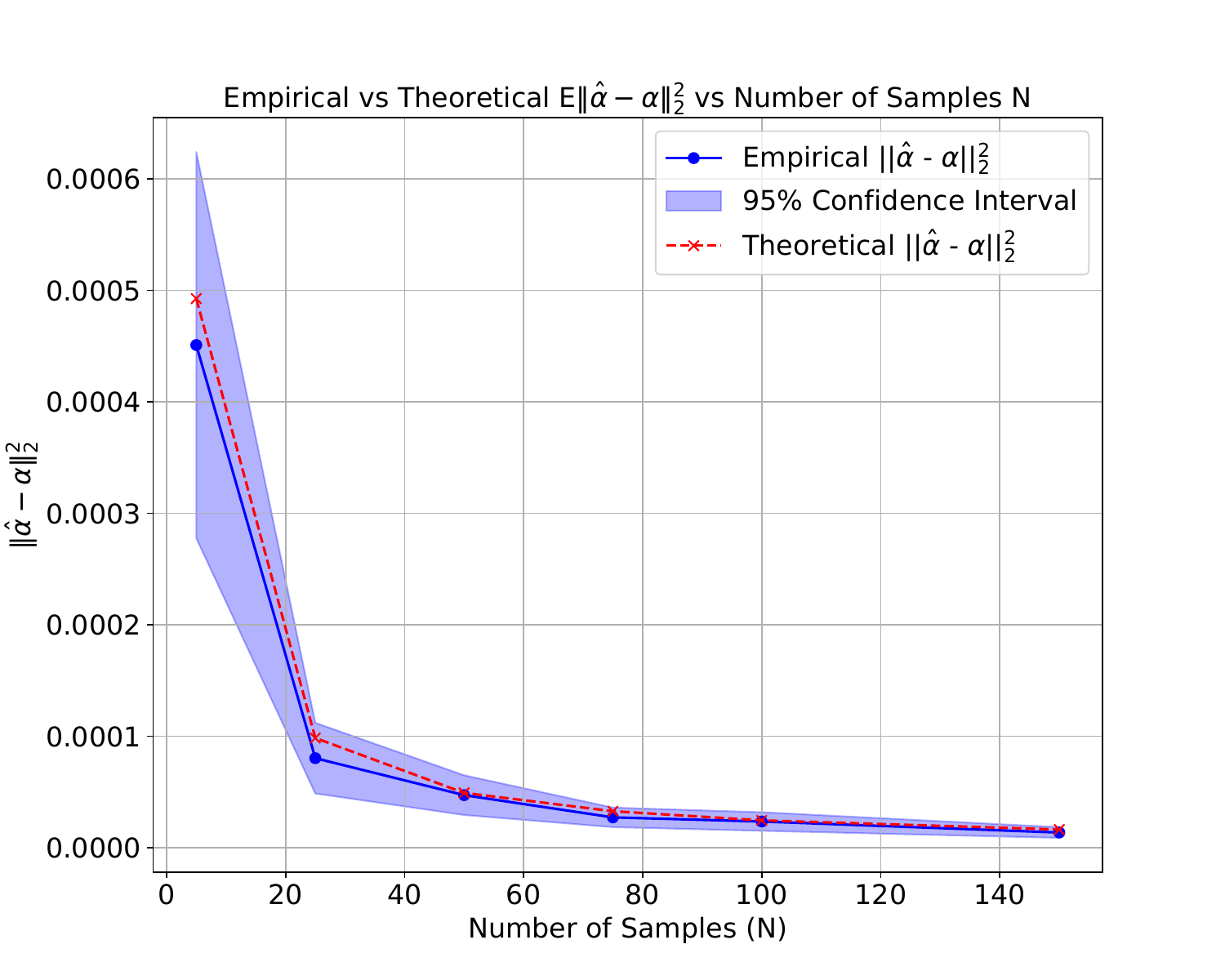}
    \caption{Simulation for Sample Complexity Analysis $\alpha = 0.8$.}
    \label{fig:sample_complexity_0.8}
\end{figure}

\begin{figure}[htbp]
    \centering
    \includegraphics[width=0.48\textwidth]{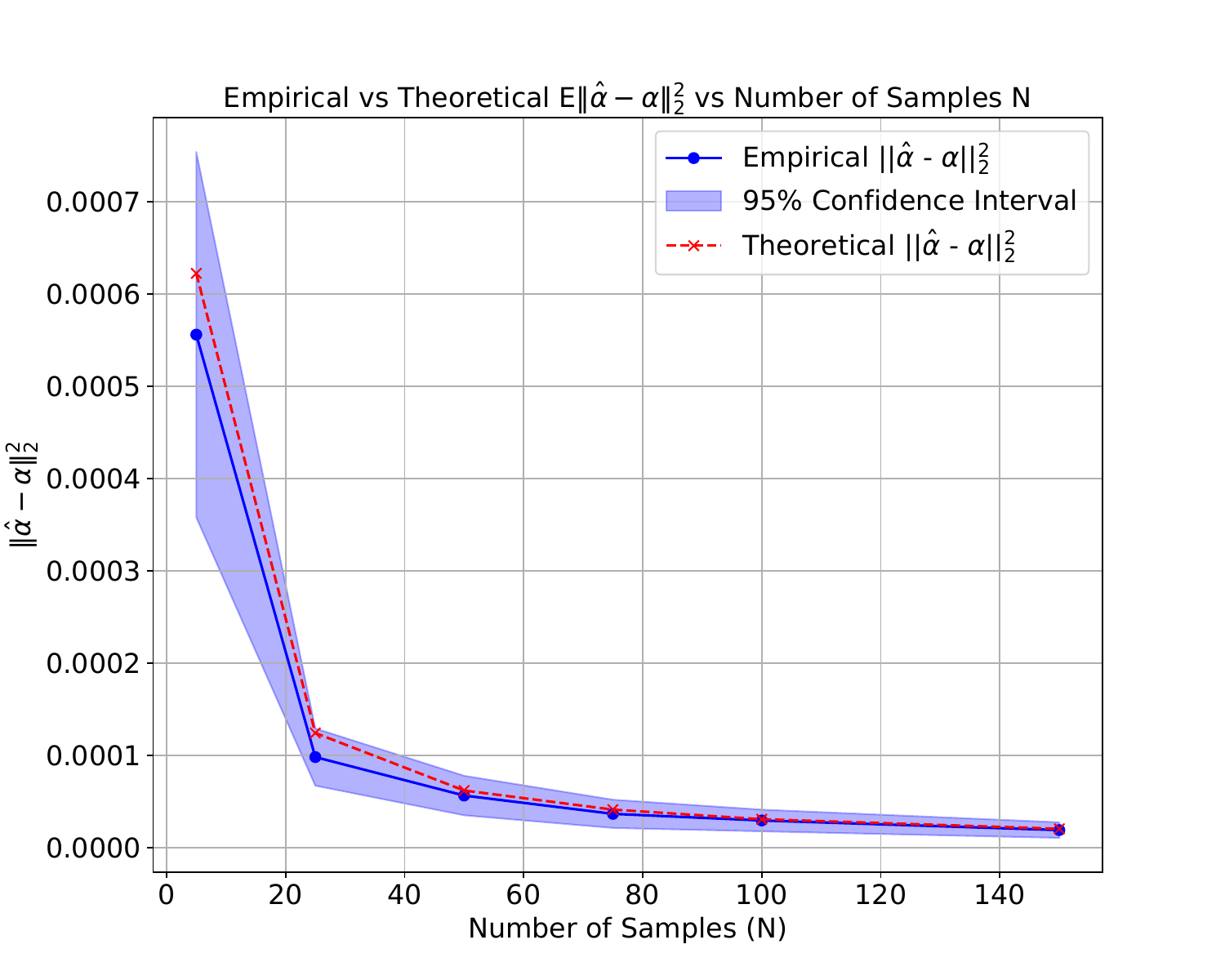}
    \caption{Simulation for Sample Complexity Analysis $\alpha = 0.6$.}
    \label{fig:sample_complexity_0.6}
\end{figure}

\begin{figure}[htbp]
    \centering
    \includegraphics[width=0.48\textwidth]{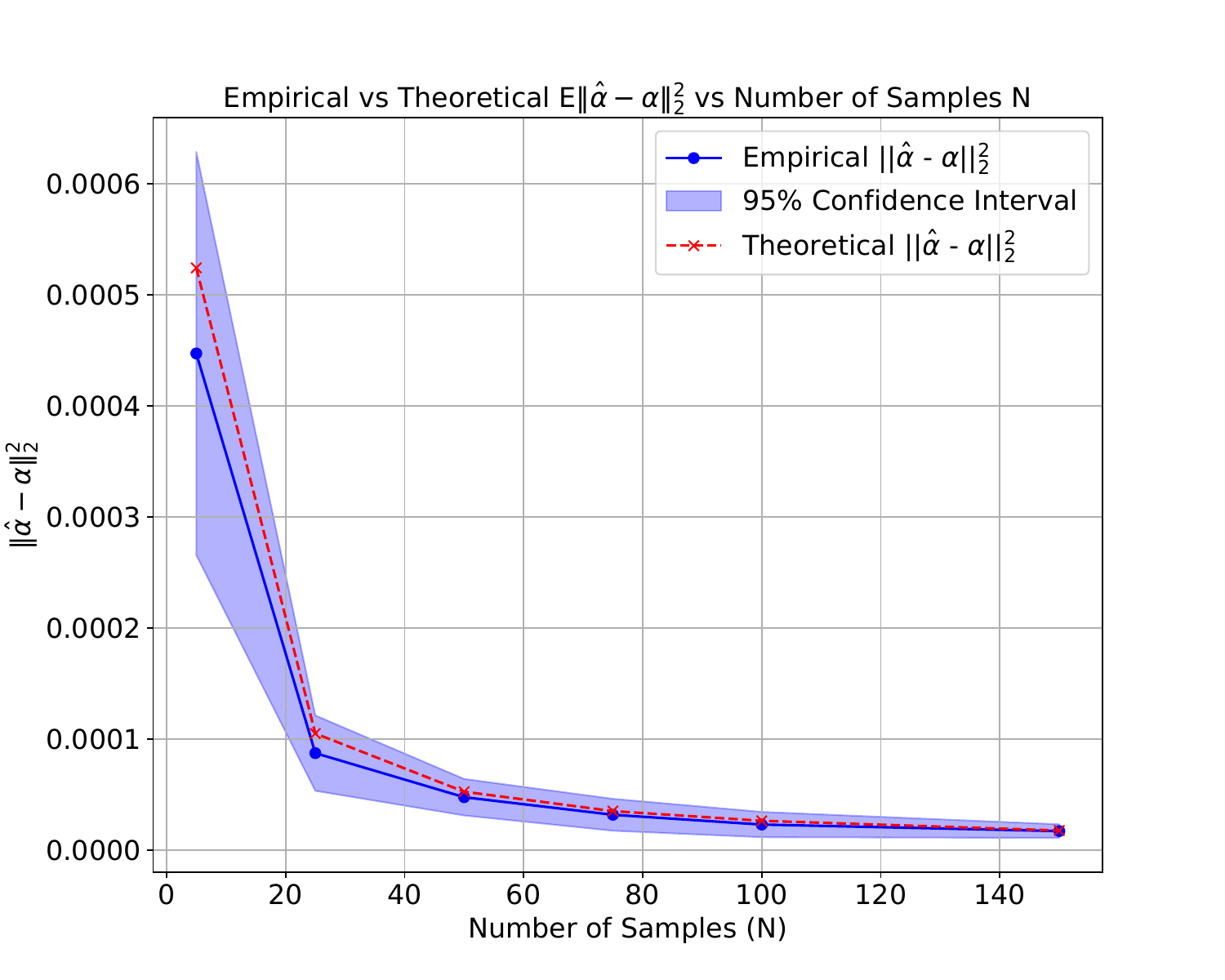}
    \caption{Simulation for Sample Complexity Analysis $\alpha = 0.4$.}
    \label{fig:sample_complexity_0.4}
\end{figure}
Figures~\ref{fig:trials_100_1}-\ref{fig:trials_100_3} show the different performance of the proposed fractional-order system identification algorithm under two assumptions. Given \( f \) and \( g \), the estimation error of the fractional order is relatively small. In the case where the functions are unknown, we set the parameters \( H = 2 \) and \( L = 7 \), and use \( L_2 \)-regularization to regularize the coefficient estimates. We achieved an highly accurate estimate of the fractional order parameter using 100 samples, specifically with \( \hat{\alpha} = 0.80702214 \) for \(\alpha = 0.8\), with \( \hat{\alpha} = 0.60713924 \) for \(\alpha = 0.6\), and with \( \hat{\alpha} = 0.40725774 \) for \(\alpha = 0.4\). We will demonstrate how the estimation error decreases as the number of samples \( N \) increases in the next part.

\subsection{Numerical Results for Sample Complexity Analysis}

We present the results of the sample complexity analysis discussed in Section IV, where \( N = O\left(\frac{d}{\epsilon}\right) \). This sample complexity bound assumes full knowledge of the functions $f$ and $g$. As shown in Figures~\ref{fig:sample_complexity_0.8}-\ref{fig:sample_complexity_0.4}, the empirical estimation error highly aligns with the theoretical error predictions.

\section{CONCLUSIONS}

In this paper, we propose an affine discrete-time fractional-order system identification algorithm. The algorithm is designed to operate under two cases: when the functions \(f\) and \(g\) are known, and when they are unknown. Our algorithm also works effectively in the regime of noisy data. The method can be summarized in two key steps. First, we generate equations for the second time step using different initial conditions. If \(f\) and \(g\) are known, the problem reduces to a least squares estimation. When \(f\) and \(g\) are unknown, we use basis function approximations to transform the problem into a least squares framework. Notably, our algorithm avoids the need to solve higher-order polynomial equations, as the generated equations depend solely on the initial conditions. This feature makes the algorithm robust in handling noisy data. Additionally, we provide both an expectation bound and a high-probability bound for the algorithm’s performance in identifying the fractional order given we know functions \(f\) and \(g\).

In future work, we will explore sample complexity analysis when the functions \(f\) and \(g\) are unknown. Selecting appropriate basis functions can be both risky and time-consuming in practice, and addressing this challenge will be a focal point of our continued research.




\section*{APPENDIX \uppercase\expandafter{\romannumeral 1\relax}}
\begin{center}
PROOF OF THEOREM 1
\end{center}

\textit{Proof.} The central idea is that the error, \( \hat{\beta} - \beta \), follows a Gaussian distribution 
\[ N\left(0, \frac{1}{N} (C^T C)^{-1} C^T K_w C (C^T C)^{-1}\right).
\]
We then apply the trace trick for the expectation, along with \textit{Lemma 1}, to derive an upper bound for 
\(
\mathbb{E}\left[\|\hat{\beta} - \beta\|_2^2\right].
\)

\textit{Lemma 1. If matrices \(A\) and \(B\) are symmetric and positive semidefinite, then 
\[
\operatorname{Trace}(AB) \leq \operatorname{Trace}(A) \lambda_{\text{max}}(B).
\]}

\textit{Proof of Lemma 1.} Since \( B \) is positive semidefinite, then
\[
\lambda_{\text{max}}(B) I - B \geq O.
\]
For any two symmetric and positive semidefinite matrices \( A \) and \( B \), we have
\[
\operatorname{Trace}(AB) \geq 0.
\]
Then we have
\[
\operatorname{Trace}\left(A\left(\lambda_{\text{max}}(B) I - B\right)\right) \geq 0, 
\]
which concludes \( 
\operatorname{Trace}(AB) \leq \operatorname{Trace}(A) \lambda_{\text{max}}(B).
\)\hfill $\square$

We consider the following optimization problem:
\[
\hat{\beta} = \arg \min_{\beta} \| Y - A \beta \|_2^2,
\]
where \( Y = [Y_1, \dots, Y_N]^T \), \( A = [C^T, C^T, \dots, C^T]^T \), and \( \beta = [\alpha_1, \dots, \alpha_d]^T \). Note that \( A \) contains \( N \) repetitions of the matrix \( C \).

The least squares solution for \( \hat{\beta} \) is given by:
\[
\hat{\beta} = (A^T A)^{-1} A^T Y
\]
which simplifies to:
\[
\hat{\beta} = (N C^T C)^{-1} (N C^T \bar{Y})
            = (C^T C)^{-1} C^T \bar{Y},
\]
where \( \bar{Y} = \frac{1}{N} \sum_{i=1}^{N} Y_i \). Similarly, the solution for \( \beta \) is:
\[
\beta = (A^T A)^{-1} A^T (Y - w)
= (C^T C)^{-1} C^T (\bar{Y} - \bar{w})
\]
where \( w = [w_1, \dots, w_N]^T \) and \( \bar{w} = \frac{1}{N} \sum_{i=1}^{N} w_i \). Thus, the error is given by:
\[
\hat{\beta} - \beta = (C^T C)^{-1} C^T \bar{w}.
\]
It is straightforward to see that:
\[
\mathbb{E}[\hat{\beta} - \beta] = 0,
\]
and the covariance is given by:
\begin{gather*}
\text{Cov}(\hat{\beta} - \beta) = (C^T C)^{-1} C^T \mathbb{E}[\bar{w} \bar{w}^T] C (C^T C)^{-1} \\
= \frac{1}{N} (C^T C)^{-1} C^T K_w C (C^T C)^{-1}.
\end{gather*}
Then, we upper bound as follows:
\begin{align*}
\mathbb{E}\left[\|\hat{\beta} - \beta\|_2^2\right] &= \mathbb{E}\left[\operatorname{Trace}\left((\hat{\beta} - \beta)^T (\hat{\beta} - \beta)\right)\right] \\
&= \mathbb{E}\left[\operatorname{Trace}\left((\hat{\beta} - \beta)(\hat{\beta} - \beta)^T\right)\right] \\
&= \operatorname{Trace}\left(\mathbb{E}\left[(\hat{\beta} - \beta)(\hat{\beta} - \beta)^T\right]\right) \\
&= \frac{1}{N} \operatorname{Trace}\left((C^T C)^{-1} C^T K_w C (C^T C)^{-1}\right) \\
&= \frac{1}{N} \operatorname{Trace}\left(C^T K_w C (C^T C)^{-2}\right) \\
&\leq \frac{1}{N} \operatorname{Trace}\left(C^T K_w C\right) \frac{1}{\lambda_{\text{min}}(C^T C)^2} \\
&= \frac{1}{N} \operatorname{Trace}\left(C C^T K_w\right) \frac{1}{\lambda_{\text{min}}(C^T C)^2} \\
&\leq \frac{1}{N} \operatorname{Trace}\left(C^T C\right) \frac{\lambda_{\text{max}}(K_w)}{\lambda_{\text{min}}(C^T C)^2} \\
&\leq \frac{d}{N} \frac{\lambda_{\text{max}}(C^T C) \lambda_{\text{max}}(K_w)}{\lambda_{\text{min}}(C^T C)^2}.
\end{align*}
The above inequalities are derived using \textit{Lemma 1}.\hfill $\square$

\section*{APPENDIX \uppercase\expandafter{\romannumeral 2\relax}}
\begin{center}
PROOF OF THEOREM 2
\end{center}

\textit{Proof.} The key idea is to whiten the error, \( \hat{\beta} - \beta \), and upper bound $\|\hat{\beta} - \beta\|_2^2$ by a chi-squared distributed random variable. We then apply known results for chi-squared distributions to derive our high-probability bound. Let $Z = \hat{\beta} - \beta \sim N(0, \Sigma)$, where \(
\Sigma = \frac{1}{N} (C^T C)^{-1} C^T K_w C (C^T C)^{-1}.
\) 
Let $V = \Sigma^{-1/2} Z \sim N(0, I)$, and we have the following:
\begin{align*}
\|Z\|_2 &= \|\Sigma^{1/2} V\|_2 \\
        &= \sqrt{V^T Q \Lambda Q^T V} \\
        &= \sqrt{W^T \Lambda W} \\
        &= \sqrt{\sum_{i=1}^{d} \lambda_i w_i^2} \\
        &\leq \sqrt{\lambda_{\text{max}}(\Sigma)} \|W\|_2 \\
        &= \sqrt{\lambda_{\text{max}}(\Sigma)} \|V\|_2.
\end{align*}
We then have:
\[
\|Z\|_2^2 \leq \lambda_{\text{max}}(\Sigma) \|V\|_2^2,
\]
where $\|V\|_2^2 \sim \chi^2(d)$, a chi-squared distribution with $d$ degrees of freedom. First, we show the proof for high-probability bound 1.

\textit{Lemma 2. 
Let $(Y_1, \dots, Y_D)$ be i.i.d. Gaussian variables with mean $0$ and variance $1$. Let $a_1, \dots, a_D$ be nonnegative. We Set:
\[
|a|_\infty = \sup_{i=1, \dots, D} |a_i|, \quad |a|_2^2 = \sum_{i=1}^{D} a_i^2,
\]
Let 
\[
Z = \sum_{i=1}^{D} a_i (Y_i^2 - 1).
\]
Then, the following inequalities hold for any positive $x$:
\[
\mathbb{P}(Z \geq 2|a|_2 \sqrt{x} + 2|a|_\infty x) \leq \exp(-x),
\]
\[
\mathbb{P}(Z \leq 2|a|_2 \sqrt{x}) \leq \exp(-x).
\]}
By \textit{Lemma 2}, The following hold for \( t > \frac{\lambda d}{2}\):
\[
\mathbb{P}(\|V\|_2^2 \geq d + 2\sqrt{d t} + 2t) \leq \exp(-t),
\]
\[
\mathbb{P}(\|Z\|_2^2 \geq \lambda \left( d + 2\sqrt{d t} + 2t \right)) \leq \exp(-t),
\]
\[
\mathbb{P}(\|Z\|_2^2 \geq t) \leq \exp\left(-\frac{\frac{t}{\lambda} + \sqrt{\frac{2td}{\lambda} - d^2}}{2}\right),
\]
where \(\lambda = \lambda_{\text{max}}(\Sigma)\). The last inequality is derived by solving for \(t\). \hfill $\square$

Next we show the proof for high-probability bound 2.

\textit{Lemma 3. Let \( Y \sim N(0,1) \), and define \( X = Y^2 \). Then, \( X \) is sub-exponential with parameters \((\nu, b) = (2, 4)\).}

\textit{Lemma 4. Suppose that \( X_1, \ldots, X_k \) are independent and \( X_k \sim \text{sub-exponential}(\nu_k, b_k) \), and \( \mathbb{E}[X_k] = \mu_k \). Then we have
\[
\mathbb{P} \left( \frac{1}{n} \sum_{i=1}^n (X_k - \mu_k) \geq t \right) \leq
\begin{cases} 
\exp \left( -\frac{nt^2}{2v_*^2} \right) & 0 \leq t \leq \frac{v_*^2}{b_*}, \\ 
\exp \left( -\frac{nt}{2b_*} \right) & t > \frac{v_*^2}{b_*}.
\end{cases}
\]
where \( b_* := \max_{k=1,\ldots,n} b_k \) and \( v_* := \sqrt{\frac{1}{n} \sum_{k=1}^n \nu_k^2} \).}

By \textit{Lemma 3} and \textit{Lemma 4}, we have:
\[
\mathbb{P} \left( \|V\|_2^2 \geq t d + d \right) \leq
\begin{cases}
\exp \left( -\frac{dt^2}{8} \right) & \text{for } 0 \leq t \leq 1, \\
\exp \left( -\frac{dt}{8} \right) & \text{for } t > 1.
\end{cases}
\]
\[
\mathbb{P} \left( \|Z\|_2^2 \geq \lambda \left( t d + d \right) \right) \leq
\begin{cases}
\exp \left( -\frac{dt^2}{8} \right) & \text{for } 0 \leq t \leq 1, \\
\exp \left( -\frac{dt}{8} \right) & \text{for } t > 1.
\end{cases}
\]
\[
\mathbb{P} \left( \|Z\|_2^2 \geq t \right) \leq
\begin{cases}
\exp \left( -\frac{d \left( \frac{t}{\lambda d} - 1 \right)^2}{8} \right) & \text{for } \lambda d \leq t \leq 2 \lambda d, \\
\exp \left( -\frac{d \left( \frac{t}{\lambda d} - 1 \right)}{8} \right) & \text{for } t > 2 \lambda d.
\end{cases}
\]
The last step is derived by substituting \( \lambda (t d + d) \) for \( t \). \hfill $\square$

\section*{CODE AVAILABILITY}

Source code is available at \url{https://github.com/EricZhang0312/Fractional_Order_System_Identification}.

\balance

\bibliography{bogdan}
\bibliographystyle{plain}

\end{document}